\documentclass[english,12pt]{article}
\usepackage[T1]{fontenc}
\usepackage[utf8]{inputenc}
\usepackage{amsmath}
\usepackage{amssymb}
\usepackage{bm}
\usepackage{color}
\usepackage{graphicx}
\usepackage[colorlinks=true,urlcolor=blue,linkcolor=blue,citecolor=red]{hyperref}
\RequirePackage[numbers,sort&compress]{natbib}
\def\be{\begin{equation}}
\def\ee{\end{equation}}
\def\ba{\begin{eqnarray}}
\def\ea{\end{eqnarray}}

\usepackage{babel}
\begin{document}

\begin{center}
{\Large \bf de Sitter invariant special relativity and}
\vskip 0.1cm 
{\Large \bf galaxy rotation curves}
\vskip 0.5cm
{\bf A. Araujo,$^{1,2}$\footnote{Email: adriana.araujo02@correo.usa.edu.co} D. F. L\'opez,$^{1,3}$\footnote{Email: difelopez@utp.edu.co} and J. G. Pereira$^1$\footnote{Email: jg.pereira@unesp.br}}
\vskip 0.3cm
$^1$ {\it Instituto de F\'{\i}sica Te\'orica, 
Universidade Estadual Paulista (UNESP) \\
%Rua Dr.~Bento Teobaldo Ferraz 271,
S\~ao Paulo, Brazil}
\vskip 0.1cm
$^2$ {\it Present address: Departamento de Matem\'aticas,
Universidad Sergio\\ Arboleda, Bogot\'a, Colombia.} 
\vskip 0.1cm
$^3$ {\it Present address: Universidad Tecnol\'ogica de Pereira \\ A.A. 97, La Julita, Pereira, Colombia.} 
\end{center}

\vskip 0.3cm
\centerline{\bf Abstract}
\begin{quote}
{\footnotesize Owing to the existence of an invariant length at the Planck scale, Einstein special relativity breaks down at that scale. A possible solution to this problem is arguably to replace the Poincar\'e invariant Einstein special relativity by a de Sitter invariant special relativity. In addition to reconciling Lorentz symmetry with the existence of an invariant length, such replacement produces concomitant changes in all relativistic theories, including general relativity, which becomes what we have called {\it de Sitter modified general relativity}. In this paper, the Newtonian limit of this theory is used to study the circular velocity of stars around the galactic center. It is shown that the de Sitter modified Newtonian force---which includes corrections coming from the underlying local kinematics---could possibly explain the rotation curve of galaxies without the necessity of supposing the existence of a dark matter halo.

}
\end{quote}

%%%%%%%%%%%%%%%%%%%%%%%%%%%%%%%%%%%%%%%%%%%%%
\section{Introduction}
\label{intro}

As quotient spaces \cite{livro}, both Minkowski and de Sitter are fundamental backgrounds for the construction of physical theories in the sense that they are known {\it a priori}, independently of Einstein equation. In particular, general relativity can be constructed on any one of them. Of course, in either case, gravitation will have the same dynamics, only the local kinematics will be different. If the underlying local spacetime is Minkowski, the local kinematics will be ruled by the Poincar\'e group of Einstein special relativity. If the underlying local spacetime is de Sitter, the local kinematics will be ruled by the de Sitter group, {\it which amounts then to replace the Poincar\'e invariant Einstein special relativity by a de Sitter invariant special relativity} \cite{dSsr0,dSsr1}. The first ideas about a de Sitter special relativity were put forward by L. Fantappi\'e, who in 1952 introduced what he called {\it Projective Relativity}, a theory further developed by G. Arcidiacono. The relevant literature can be traced back from Ref.~\cite{FA}.

As a homogeneous space, the de Sitter spacetime has the constant scalar curvature
\be
R = 12 \, l^{-2} \, ,
\label{RicciScalar}
\ee
where $l$ is the de Sitter length-parameter, which is related to the cosmological term through $\Lambda = 3/l^2$. By definition, Lorentz transformations do not change the curvature of the homogeneous spacetime in which they are performed. From Eq.~(\ref{RicciScalar}), we see that {\it Lorentz transformations leave the de Sitter length parameter $l$ invariant} \cite{ccc}. Although somewhat hidden in Minkowski spacetime, because what is left invariant in this case is an infinite length---corresponding to a vanishing scalar curvature---in de Sitter spacetime, whose pseudo-radius is finite, this property becomes manifest. Contrary to the usual belief, therefore, {\it Lorentz transformations do leave invariant a very particular length parameter: that defining the scalar curvature of the homogeneous spacetime in which they are performed}. If the Planck length $l_P$ is to be invariant under Lorentz transformations, it must then represent the pseudo-radius of spacetime at the Planck scale, which will be a de Sitter space with the Planck cosmological term
\be
\Lambda_P = 3 / l_P^{2} \simeq 1.2 \times 10^{70}\, {\rm m}^{-2}.
\label{PlanLam}
\ee
In de Sitter invariant special relativity, therefore, the existence of an invariant length-parameter does not clash with Lorentz invariance, which remains a symmetry at all scales.

Through a simple algebraic manipulation, expression (\ref{PlanLam}) can be rewritten in the form
\begin{equation}
\Lambda_P = \frac{4 \pi G}{c^4} \, \varepsilon_P  \, ,
\label{kineLambda0}
\end{equation}
where $\varepsilon_P$ is the Planck energy density. Now, this expression can be considered an extremal case of a general expression relating the local cosmological term to the corresponding energy density of a physical system. Accordingly, a physical system with energy density $\varepsilon_m$ will induce the local cosmological term \cite{LocaLambda}
\begin{equation}
\Lambda = \frac{4 \pi G}{c^4} \, \varepsilon_m
\label{kineLambda}
\end{equation}
in the region occupied by the system. The idea that a physical system with energy density $\varepsilon_m$ could induce a local cosmological term $\Lambda$ in the underlying local spacetime was first considered by Mansouri \cite{mansouri}. Such a change in the local structure of spacetime is necessary to comply with the local kinematics, now governed by the de Sitter group. It should be noted that this local cosmological term $\Lambda$ is different from the usual notion in the sense that it is no longer required to be constant \cite{hendrik}. For example, outside the region occupied by the physical system, where $\varepsilon_m$ vanishes, the cosmological term $\Lambda$ vanishes as well. In a sense, it can be thought of as an asymptotically flat de Sitter spacetime.

As an illustration, let us recall that the space section of spacetime is nearly flat today. This means that the mean energy density of the Universe is of the same order of the Hubble critical energy density \cite{Hubble}
\begin{equation}
\varepsilon_m \simeq 10^{-9}~{\rm Kg} \, {\rm m}^{-1} \, {\rm s}^{-2} \, .
\end{equation}
Using this value in expression (\ref{kineLambda}), the effective cosmological term of the present-day universe is found to be
\begin{equation}
\Lambda \simeq 10^{-52}~\mbox{m}^{-2} \, ,
\label{LambdaToday}
\end{equation}
which is the order of magnitude of the observed value \cite{obs1,obs2,obs3}. It is important to remark that, according to the de Sitter invariant special relativity, the cosmological term $\Lambda$ shows up as a kinematic effect. This means that $\Lambda$ is not a dynamical variable, but an external parameter standing on an equal footing with the energy density $\varepsilon_m$ of the physical system.

When general relativity is constructed on a de Sitter background, all solutions to Einstein equation turn out to be a spacetime that reduces locally to de Sitter. In this case, general relativity changes to what we have called {\it de Sitter modified general relativity}. By considering this theory, we have already obtained its Newtonian limit, as well as the Newtonian Friedmann equations \cite{paper1}. The purpose of the present paper is to use the same Newtonian limit to study galaxy rotation curves. For the sake of completeness, we review in the next section the de Sitter modified Einstein equation and its Newtonian limit.

%%%%%%%%%%%%%%%%%%%%%%%%%%%%%%%%%%%%
\section{The de Sitter modified Newtonian potential}
\label{dSitterGR}

In de Sitter modified general relativity, the kinematic curvature of the underlying de Sitter spacetime and the dynamical curvature of general relativity are both included in the same Riemann tensor \cite{CartanGeo}. As a consequence, the cosmological term $\Lambda$ does not explicitly appears in Einstein's equation, and the second Bianchi identity does not require it to be constant \cite{ccc,hendrik}. Far away from the Planck scale, $\Lambda$ can consequently assume smaller values, corresponding to larger values of the de Sitter length-parameter $l$. For low energy systems, like for example the present-day universe, the value of $\Lambda$ will be very small, and the de Sitter invariant special relativity will approach the Poincar\'e invariant Einstein special relativity.

%%%%%%%%%%%%%%%%%%%%%%%%%%%%%%%%%%%%
\subsection{The de Sitter modified Einstein equation}
\label{dSmodi10}

In a locally de Sitter spacetime, the gravitational action is written in the form 
\begin{equation}
S_g = - \frac{c^3}{16 \pi G} \int R \sqrt{-g} \, d^4x \,
\end{equation}
where the scalar curvature $R$ represents both the {\it kinematical curvature} of the underlying de Sitter spacetime and the {\it dynamical curvature} of general relativity. Variation of the gravitational action under a general metric shift $\delta g_{\rho \mu}$ yields
\begin{equation}
\delta S_g = \frac{c^3}{16 \pi G} \int \big({R}^{\rho \mu} - {\textstyle{\frac{1}{2}}} g^{\rho \mu} {R} \big)
\delta g_{\rho \mu} \sqrt{-g} \, d^4x \, .
\label{AgVari}
\end{equation}
The term between parentheses is the so-called Einstein tensor, which satisfies the contracted form of the second Bianchi identity
\begin{equation}
\nabla_\mu \big({R}^{\rho \mu} - {\textstyle{\frac{1}{2}}} g^{\rho \mu} {R} \big) = 0 \, ,
\label{2BiId}
\end{equation}
with $\nabla_\mu$ the covariant derivative in the Levi-Civita connection of the spacetime metric.

Let us consider now the action of a general source (or matter) field
\begin{equation}
S_m = \frac{1}{c} \int {\mathcal L}_m \, \sqrt{-g} \, d^4x \, .
\label{A}
\end{equation}
Variation of this action under a general metric shift $\delta g_{\rho \mu}$ yields
\begin{equation}
\delta S_m = - \frac{1}{2 c} \int T^{\rho \mu} \, \delta g_{\rho \mu} \, \sqrt{-g} \, d^4x \, ,
\label{deltaA00}
\end{equation}
where 
\begin{equation}
T^{\rho \mu} = -\, \frac{2}{\sqrt{-g}} \, \frac{\delta (\sqrt{-g} \, {\mathcal L}_m)}{\delta g_{\rho \mu}}
\label{syem0}
\end{equation}
is the symmetric energy-momentum tensor. When considering variations coming from a general coordinate transformation, however, it is necessary to take into account some subtleties. The problem is that the explicit form of the covariantly conserved current depends on the local properties of spacetime. The metric transformation in this case must take into account those properties. For example, in the case of ordinary general relativity, whose spacetime reduces locally to Minkowski, the action increment assumes the form
\begin{equation}
\delta S_m = \frac{1}{c} \int \delta^\rho_\alpha \, T^{\alpha \mu} \, \nabla_\rho \varepsilon_\mu(x) \, \sqrt{-g} \, d^4x \, ,
\label{deltaA000}
\end{equation}
where $\delta^\rho_\nu$ are the Killing vectors of translations---the transformations that define the transitivity of Minkowski spacetime---and $\varepsilon_\mu(x)$ are the transformation parameters. Invariance of $S_m$ yields the conservation law
\begin{equation}
\nabla_\mu \big(\delta^\rho_\alpha \, T^{\alpha \mu} \big) = 0 \, .
\label{UniCon0}
\end{equation} 

We have on purpose kept the translational Killing vectors $\delta^\rho_\alpha$ in the above expressions because they are quite elucidative. For example, remember that Noether's theorem establishes a relation between invariance under ordinary translations and energy-momentum conservation. The presence of the translational Killing vectors in the conserved current leaves it clear that the conservation law (\ref{UniCon0}) is a natural consequence of Noether's theorem, in the sense that the local properties of spacetime were properly taken into account.\footnote{In the usual formulation of general relativity, the translational Killing vectors are not explicitly shown in the conserved quantities. Although this can be done---because the Killing vectors are just Kronecker delta's---it becomes unclear why the invariance of the Lagrangian under diffeomorphism should give the energy-momentum conservation, a current whose conservation law is related, through Noether's theorem, to the invariance of the source Lagrangian under spacetime translations.}

Let us consider now the case of locally de Sitter spacetimes. Analogously to (\ref{deltaA000}), the variation of $S_m$ under a metric shift that takes into account the locally de Sitter property of spacetime is
\begin{equation}
\delta S_m = \frac{1}{c} \int \xi^{(\rho}_\alpha \, T^{\alpha \mu)} \, \nabla_\rho \varepsilon_\mu(x) \, \sqrt{-g} \, d^4x \, ,
\label{deltaA0}
\end{equation}
where $\xi^\rho_\nu$ are the Killing vectors of the de Sitter ``translations'', the transformations that define the transitivity of de Sitter spacetime \cite{gursey}. Invariance of $S_m$ yields the conservation law
\begin{equation}
\nabla_\mu \Pi^{(\rho \mu)} = 0 \, ,
\label{UniCon}
\end{equation}
with $\Pi^{(\rho \mu)}$ the symmetric part of the current \cite{dSgeod}
\begin{equation}
\Pi^{\rho \mu} = \xi^{\rho}_{\alpha} \, T^{\alpha \mu} \, .
\label{TmK}
\end{equation}

In locally de Sitter spacetimes, therefore, the action variation (\ref{deltaA00}) assumes the form
\begin{equation}
\delta S_m = - \frac{1}{2 c} \int \Pi^{(\rho \mu)} \, \delta g_{\rho \mu} \, \sqrt{-g} \, d^4x \, .
\label{deltaA2}
\end{equation}
Thus, from the variational principle $\delta S_g + \delta S_m = 0$, we get
\begin{equation}
\frac{c^3}{16 \pi G} \int \Big({R}^{\rho \mu} - \frac{1}{2} g^{\rho \mu} {R} -
\frac{8 \pi G}{c^4} \, \Pi^{(\rho \mu)} \Big)
\delta g_{\rho \mu} \sqrt{-g} \; d^4x = 0 \, .
\end{equation}
In view of the arbitrariness of $\delta g_{\rho \mu}$, the de Sitter modified Einstein equation is found to be
\begin{equation}
{R}^{\rho \mu} - \frac{1}{2} g^{\rho \mu} {R} =
\frac{8 \pi G}{c^4} \, \Pi^{(\rho \mu)} \, .
\label{NewEinstein}
\end{equation}
This is the equation that replaces ordinary Einstein equation when the Poin\-ca\-r\'e invariant Einstein special relativity is replaced by a de Sitter-invariant special relativity \cite{livro,dSsr0,dSsr1}. In the contraction limit $l \to \infty$, which corresponds to $\Lambda \to 0$, the underlying de Sitter spacetime contracts to Minkowski, the de Sitter Killing vectors $\xi^\mu_\rho$ reduce to the Killing vectors $\delta^\mu_\rho$ of ordinary translations, and we recover the ordinary Einstein equation
\begin{equation}
R^{\rho \mu} - {{\textstyle{\frac{1}{2}}}} g^{\rho \mu} R = 
\frac{8 \pi G}{c^4} \; T^{\rho \mu}
\label{OldEinstein}
\end{equation}
for locally Minkowski spacetimes.

%%%%%%%%%%%%%%%%%%%%%%%%%%%%%%%%%%%%
\subsection{The de Sitter modified Newtonian limit}

In a locally de Sitter spacetime, and in the weak field approximation, the spacetime metric is expanded according to
\begin{equation}
g_{\mu\nu} = \hat{g}_{\mu\nu} + h_{\mu\nu} \, ,
\end{equation}
where $\hat{g}_{\mu\nu}$ represents the background de Sitter metric and $h_{\mu\nu}$ is the gravitational perturbation. The background connection, which corresponds to the zeroth-order connection, is
\begin{equation}
\hat \Gamma^{\rho}{}_{\mu \nu} = {\textstyle{\frac{1}{2}}} \, \hat{g}^{\rho \lambda}
\big(\partial_\mu \hat{g}_{\lambda \nu} + \partial_\nu \hat{g}_{\mu \lambda} -
\partial_\lambda \hat{g}_{\mu \nu} \big).
\label{0conn}
\end{equation}
The corresponding Riemann tensor $\hat{R}^\alpha{}_{\beta \mu \nu}$ represents the curvature of the non-gravitational de Sitter background. In harmonic coordinates, which is expressed by the condition
\begin{equation}
\hat \nabla_{\nu}h^{\rho\nu} - {\textstyle{\frac{1}{2}}} \hat \nabla^{\rho}h = 0 \, ,
\label{armonico}
\end{equation}
with $h = h^\mu{}_\mu$, the first-order Ricci tensor is found to be
\begin{equation}
R^{\mbox{\tiny{(1)}}}_{\mu\nu} = - \, {\textstyle{\frac{1}{2}}} \hat \Box h_{\mu \nu} +
h^\sigma{}_{(\nu} \, \hat R_{\sigma \mu)} -
h^\rho{}_\sigma \, \hat R^\sigma{}_{(\mu \rho \nu)}.
\label{Ricci1bis}
\end{equation}
The de Sitter modified Einstein equations are then given by
\begin{equation}
- \, {\textstyle{\frac{1}{2}}} \hat \Box h_{\mu \nu} +
h^\sigma{}_{(\nu} \, \hat R_{\sigma \mu)} -
h^\rho{}_\sigma \, \hat R^\sigma{}_{(\mu \rho \nu)} = \frac{8 \pi G}{c^4} \Big( \Pi_{\mu \nu} -
{\textstyle{\frac{1}{2}}} \, \hat g_{\mu \nu} \Pi \Big) \, .
\label{EEbis}
\end{equation}

The usual Newtonian limit is obtained when the gravitational field is weak and the particle velocities are small. In the presence of a cosmological term $\Lambda$, however, it has some additional subtleties. In the process of group contractions, the Galilei group is obtained from Poincar\'e in the contraction limit $c \to \infty$. The Newton-Hooke group, on the other hand, does not follow straightforwardly from the de Sitter group through the same limit. The reason is that, under such limit, the boost transformations are lost. In order to obtain a physically acceptable result, one has to simultaneously consider the limits $c \to \infty$ and $\Lambda \to 0$, but in such a way that $c^2 \Lambda = \tau^{-2}$, with $\tau$ a time parameter. This means that the usual weak field condition of Newtonian gravity must be supplemented with the small $\Lambda$ condition \cite{gibbons}
\begin{equation}
\Lambda r^2 \ll 1\, .
\end{equation}
In this limit, and identifying
\begin{equation}
h_{00} = 2 \phi/c^2,
\end{equation}
with $\phi$ the gravitational scalar potential, the de Sitter modified Einstein equation (\ref{EEbis}) assumes the form
\begin{equation}
\hat \Delta \phi +
2 \phi \, \hat R_{00} = \frac{4 \pi G}{c^2} \, \Pi_{00} \, ,
\label{EE3}
\end{equation}
where $\hat \Delta$ is the Laplace operator in the background de Sitter metric $\hat{g}^{i j}$, and
\begin{equation}
\Pi_{00} = \xi^0_0 \, T_{00} \, ,
\label{pixite}
\end{equation}
with $\xi^0_0$ the zero-component of the Killing vectors of the de Sitter ``translations'' and $T_{00} = \rho c^2$. In static coordinates, in which we can write $\Pi_{00} = \rho_\Pi c^2$, with
\begin{equation}
\rho_\Pi \simeq \rho \left(1 - r^2/2 l^2 \right) \, ,
\label{rhoPi}
\end{equation}
the solution of equation (\ref{EE3}) yields the de Sitter modified Newtonian potential \cite{paper1}
\begin{equation}
\phi(r) = 
- \frac{G M}{r} - \frac{G M \Lambda}{6} \, r \, ,
\label{ModNewton0}
\end{equation}
where we have used the relation $\Lambda = 3/l^2$. In the limit of a vanishing cosmological term $\Lambda \to 0$, it reduces to the ordinary Newtonian potential. It is important to reinforce that, since the role of the de Sitter group is to govern the underlying local kinematics, $\Lambda$ shows up not as part of dynamics, but as a kinematic effect.

%%%%%%%%%%%%%%%%%%%%%%%%%%%%%%%%%%%%
\section{Galaxy rotation curves}

\begin{figure}[http]
\begin{center}
\scalebox{0.58}{\includegraphics{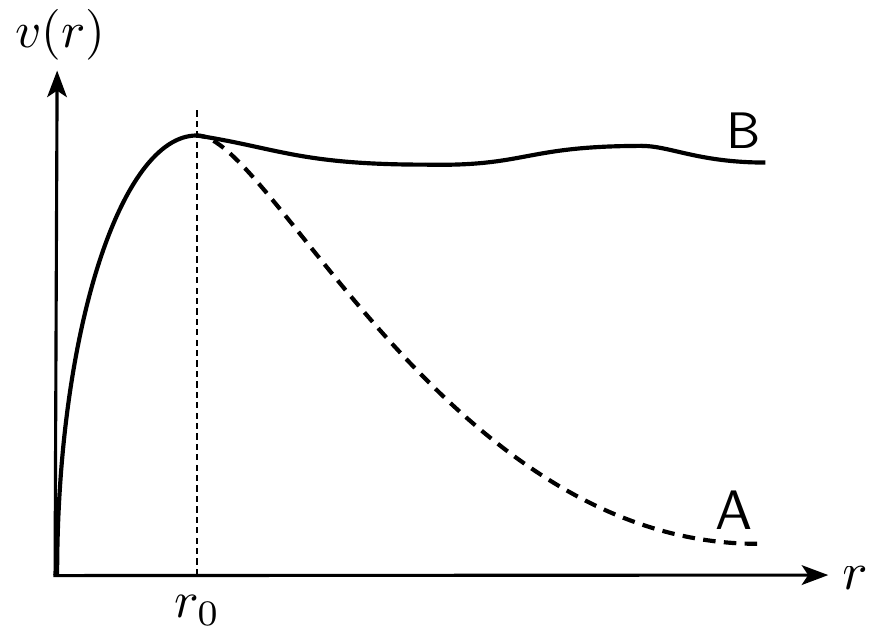}}
\caption{\it A typical galaxy rotation curve.} 
\label{Fig1}
\end{center}
\end{figure}

A generic galaxy rotation curve can be divided into three regions: (i) an inner region (aka bulge) in which the rotation velocity of the stars rises linearly with the distance from the center; (ii) a region where the speed reaches a maximum and then begins to decrease (at the so-called {turn over radius}); (iii) and a Keplerian region in which the whole mass of the galaxy can be assumed to be given by the bulge mass $M_0$, and located at the central point. For this reason, the Newtonian gravitational force in the Keplerian region resembles that of a point mass force,
\begin{equation}
F = - \frac{G M_0}{r^2} \, .
\end{equation}
The corresponding rotation velocity of galaxies is found to be \cite{GalaDyn}
\begin{equation}
v(r) \equiv \sqrt{r \, |F(r)|} = \sqrt{{G M_0}/{r}} \, ,
\label{Newtvel}
\end{equation}
from where we see that the velocity falls off as $v \sim r^{-1/2}$, as schematically depicted in curve {\sf A} of Figure~\ref{Fig1}. However, instead of such behavior, galaxies show in general a flat rotation curve, as depicted in curve {\sf B} of Figure~\ref{Fig1}. We can then say that either there is some unaccounted matter in the galaxy---usually called dark matter---or gravity behaves different from the usual Newtonian limit.
%%%%%%%%%%%%%%% FIGURE %%%%%%%%%%%%%
%%%%%%%%%%%%%%%%%

In addition to the traditional approach that takes for granted the existence of dark matter \cite{dm}, there are some alternative attempts to explain the original missing mass problem in galaxies and cluster of galaxies. These alternative approaches assume that an appropriate modification of gravitation could conceivably eliminate the need (either partially or totally) for dark matter. The most well-known theory of this kind is MOND, which changes the dynamics of gravity at the non-relativistic level \cite{mond}. Another example that is worth mentioning is a modified gravity theory based on entropic gravity \cite{verlinde}. An alternative model, which like the approach described in this paper is also based on the Fantappi\'e-Arcidiacono projective special relativity, is the so-called Chiatti-Licata projective general relativity \cite{Chiatti}. According to this theory, possible local violations of the inertia principle---induced by the quantized granular structure of the cosmic inertial field---could produce physical effects that mimics exactly those associated to ordinary dark matter. In the present paper, we study another modified-gravity theory that is also grounded on the Fantappi\'e-Arcidiacono projective special relativity. Differently from the Chiatti-Licata model, however, the corrections to ordinary general relativity are not related to a violation of the inertial principle, but to a change in the local kinematics of spacetime.

%%%%%%%%%%%%%%%%%%%%%%%%%%%%%%%%%%%%%%%%%%
\subsection{The de Sitter special relativity and galaxy rotation curves}

We consider now the de Sitter modified Einstein equation, obtained in Section~\ref{dSmodi10}, whose Newtonian limit has already been shown to yield the gravitational potential (\ref{ModNewton0}), with $\Lambda$ given by Eq.~(\ref{kineLambda}). Considering that the galaxy mass density $\rho_m$ is not constant, $\Lambda$ is not constant as well, and the corresponding gravitational force $F = - d \phi(r)/d r$ assumes the form
\begin{equation}
F = - \frac{G M}{r^2} + \frac{G M \Lambda(r)}{6} + 
\frac{G M}{6} \, r \frac{d \Lambda(r)}{d r} \, .
\label{ModNewtForceBis}
\end{equation}
The first term on the right-hand side represents the usual attractive Newtonian force. The background de Sitter spacetime contributes with an additional repulsive force proportional to $\Lambda(r)$, as well as with a force proportional to the radial derivative of $\Lambda(r)$, which will be attractive or repulsive depending on the sign of $d \Lambda(r)/dr$. In what follows we are going to use the gravitational force (\ref{ModNewtForceBis}) to study the circular velocity of a star around the galactic center.

Let us begin with the inner region of the galaxy $r \ll r_0$, where the circular velocity of the stars rises almost linearly with $r$. This means that in this region only the Newtonian force is in action, and the mass density $\rho_m(r)$ of the galaxy decreases slowly with the radius $r$. In fact, for a nearly constant mass density $\rho_m(r) \simeq \rho_0$, the inner mass $M(r)$ assumes the form $M(r) = (4/3) \pi \rho_0 r^3$, and the Newtonian star velocity is easily seen to grow linearly with $r$
\begin{equation}
v(r) = \sqrt{(4/3) \pi G \rho_0} \; r \, ,
\end{equation}
in agreement with observations. On the other hand, in the Keplerian region, defined by $r \gg r_0$, the Newtonian force becomes negligible and the relevant force takes the form
\begin{equation}
F = \frac{G M_0}{6} \, \Lambda(r) + \frac{G M_0}{6} \, r \frac{d \Lambda(r)}{d r} \, ,
\label{ModNewtForceBis3}
\end{equation}
where we have assumed that in this region the whole mass of the galaxy can be represented by the bulge mass $M_0$. The squared circular velocity of a star at a distance $r$ from the galactic center is now given by
\begin{equation}
v^2(r) \equiv r |F(r) | = \frac{GM_0}{6} \bigg[\Lambda(r) r + 
r^2 \frac{d\Lambda(r)}{dr} \bigg] \, .
\label{SqVelCons}
\end{equation}

Now, the above expression has a solution in which $v^2(r)$ is constant. Such solution is obtained when $\Lambda(r)$ satisfies the first-order differential equation
\begin{equation}
r^2 \, \frac{d \Lambda(r)}{dr} + \Lambda(r) r = \beta \, ,
\label{DiffEq1}
\end{equation}
with $\beta$ a constant.
It is convenient at this point to introduce the dimensionless coordinate $r' = r/r_0$, in terms of which the differential equation (\ref{DiffEq1}) assumes the form
\begin{equation}
r'^2 \, \frac{d \Lambda(r)}{dr'} + \Lambda(r) r' - \frac{\beta}{r_0} = 0 \, .
\label{DiffEq2}
\end{equation}
In terms of the original variables, its solution is
\begin{equation}
\Lambda(r) = \frac{\beta}{r} \, \ln\bigg(\frac{r}{r_0} \bigg) +
\gamma \, \frac{r_0}{r} \, ,
\end{equation}
where $\gamma$ is an integration constant. Since at $r=r_0$ the cosmological term has the value $\Lambda(r) = \Lambda_0$, we can immediately infer that $\gamma = \Lambda_0$. Imposing furthermore that $d \Lambda(r) / dr = 0$ at $r = r_0$, we find that $\beta = \Lambda_0 r_0$. The final form of the solution is consequently
\begin{equation}
\Lambda(r) = \Lambda_0 \left[ \frac{r_0}{r} \, \ln\bigg(\frac{r}{r_0}\bigg) +
\frac{r_0}{r}  \right] \, .
\end{equation}
On account of the relation (\ref{kineLambda}), in terms of the mass density the solution is written as
\begin{equation}
\rho(r) = \rho_0 \left[ \frac{r_0}{r} \, \ln\bigg(\frac{r}{r_0}\bigg) +
\frac{r_0}{r}  \right] \, .
\label{MagicMDP}
\end{equation}
The combination of this fiducial mass density profile $\rho(r)$ with the de Sitter modified Newtonian force (\ref{ModNewtForceBis}) naturally yields a flat rotation curve for the galaxy, without necessity of supposing the existence of a dark matter halo. It should be remarked that the mass density profile (\ref{MagicMDP}) represents a small correction to the power law $\rho(r) \simeq \rho_0 (r_0/r)$, which is within the class of physically acceptable profiles \cite{GalaDyn}.

We turn now to the question of the order of magnitude of the circular velocity of the stars around the galactic center. According to Eqs.~(\ref{SqVelCons}) and (\ref{DiffEq1}), the squared velocity of the flat portion of a galaxy rotation curve is given by
\begin{equation}
v_0^2 = \frac{G M_0}{6} \, \beta \equiv \frac{G M_0}{6} \, \Lambda_0 r_0 \, ,
\label{Vflat}
\end{equation}
where, we recall, the subscript zero refers to the values at the turn over region---the transition region from the bulge to the disk of the galaxy. On the other hand, astronomical observations show that the squared velocity of the flat portion of a galaxy rotation curve is of the order
\begin{equation}
v_{\rm obs} \simeq 10^{10}~{\rm m}^2 {\rm s}^{-2} \, .
\label{obsV}
\end{equation}
Using the Milky Way values for the mass and radius of a typical galaxy bulge, given respectively by \cite{MassDens}
\[
M_0 = 10^{10} M_{\odot} \simeq 2 \times 10^{40}~{\rm Kg} \qquad {\rm and} \qquad
r_0 = 3~{\rm kpc} \simeq 10^{20}~{\rm m} \, ,
\]
the squared velocity~(\ref{Vflat}) will coincide with the observed value (\ref{obsV}) provided the cosmological term $\Lambda_0$ in the turn over region is of the order 
\begin{equation}
\Lambda_0 \equiv \frac{4 \pi G}{c^2} \rho_0 \simeq 10^{-40}~{\rm m}^{-2} \, ,
\label{L0}
\end{equation}
which is equivalent to a mass density of the order
\begin{equation}
\rho_0 \simeq 10^{-14}~{\rm Kg} \, {\rm m}^{-3} \, .
\label{rhoZero}
\end{equation}

For the sake of comparison, let us note that the typical mass density of a galactic nucleus, a core region of the bulge with radius $r_c \simeq 1~{\rm pc}$, is estimated to be (see Ref.~\cite{GalaDyn}, page 29)
\begin{equation}
\rho_c \simeq 5 \times 10^6 M_{\odot} \, {\rm pc}^{-3} = 3 \times 10^{-13}~{\rm Kg \, m^{-3}} \, .
\label{rhoCentral}
\end{equation}
Considering that the mass density of the bulge is nearly constant, or at most falls off slowly with the radius, the mass density $\rho_0$ at the distance $r_0$ from the galactic center, given by Eq.~(\ref{rhoCentral}), is physically reasonable in the sense that it is just one order of magnitude smaller then the density $\rho_c$ at the central region of the bulge. This is an important constraint, which is crucial for the correct description of the galaxy rotation curve.

%%%%%%%%%%%%%%%%%%%%%%%%%%%%%%%%%%%
\section{Final remarks}
\label{Fr}

Due to the existence of an invariant length at the Planck scale, which is not allowed by ordinary special relativity, there is a widespread belief that Lorentz symmetry should break down at that scale. However, this is not necessarily true. In fact, if one replaces the Poincar\'e invariant Einstein special relativity by a de Sitter invariant special relativity, it is possible to reconcile Lorentz symmetry with the existence of an invariant length, not only at the Planck scale but at all energy scales. On the other hand, the replacement of ordinary special relativity by a de Sitter invariant special relativity produces concomitant changes in all relativistic theories, including general relativity, giving rise to what we have called {\it de Sitter modified general relativity}.
In a recent companion paper \cite{paper1}, we have obtained the Newtonian limit of this theory, as well as the Newtonian Friedmann equations. Using these results, we have shown that such theory is able to give a reasonable account of the dark energy content of the present-day universe.

In the present paper, we have used the same Newtonian limit to study galaxy rotation curves. The main difference of the de Sitter modified Newtonian limit in relation to the usual one is the existence of two new {\it kinematic} forces, as can be seen from Eq.~(\ref{ModNewtForceBis}). The first one is repulsive and is proportional to $\Lambda$, whereas the second is proportional to the radial derivative of the cosmological term $\Lambda$. Since $d \Lambda / dr < 0$ in the galactic disk, this new force is {\it attractive}. Most importantly, it vanishes in the galactic bulge, becoming active only in the Keplerian region of the galaxy, where $\Lambda$ decays faster. Using the de Sitter modified Newtonian force, we have obtained a kind of fiducial mass density profile, given by Eq.~(\ref{MagicMDP}), which yields a flat rotation curve for the galaxy without the necessity of supposing the existence of a dark halo. It is important to remark that this fiducial mass density profile is within the class of physically acceptable profiles \cite{GalaDyn}.

Of course, not all galaxies show a perfectly flat rotation curve. In spite of this fact, these galaxies can still be studied in the present context: one has simply to replace Eq.~(\ref{DiffEq1}) by
\begin{equation}
r^2 \, \frac{d \Lambda(r)}{dr} + \Lambda(r) r = \beta(r) \, ,
\label{DiffEq1Bis}
\end{equation}
with $\beta(r)$ a function describing the behavior of the galaxy rotation curve in the Keplerian region. The solution to this equation is easily found to be
\begin{equation}
\Lambda(r) = \frac{1}{r} \int \frac{1}{r} \, \beta(r) dr + 
\gamma \, \frac{r_0}{r} \, .
\label{solution2}
\end{equation}
Considering that the explicit form of $\beta(r)$ can be inferred from observations, one can then find the explicit form of $\Lambda(r)$, or equivalently, the explicit form of the mass density profile $\rho(r)$ that gives rise to the observed rotation curve. Conversely, given a specific mass density profile, we can proceed backward to find $\beta(r)$, which determines the corresponding galaxy rotation curve through Eq.~(\ref{SqVelCons}). This means that this theory can be applied individually for each galaxy, taking into account their different specificities. In other words, the theory can be experimentally tested for each individual galaxy. 

In view of recent results favoring a gravitational solution to the missing mass problem \cite{NoDM5,NoDM6}, as well as of the lack of experimental sign of particles that could play the role of dark matter \cite{noDM1,noDM2,noDM3,noDM4}, the approach discussed in this paper may eventually constitute an alternative gravitational paradigm for the study of the missing mass problem.

%%%%%%%%%%%%%%%%%%%%%%%%%%%%%%%%%%%%
\section*{Acknowledgments}

AA and DFL would like to thank CAPES/Brazil (33015015001P7) for scholarship grants. JGP thanks CNPq/Brazil (304190/2017-9) for a research grant.

%%%%%%%%%%%%%%%%%%%%%%%%%%%%%%%%%%%%%%%%%%

\end{document}